\documentclass[8pt,preprintnumbers,amsmath,amssymb,prl,twocolumn]{revtex4}
\usepackage{graphicx}
\usepackage{dcolumn}
\usepackage{bm}
\usepackage[colorlinks=true,linkcolor=blue,anchorcolor=blue,citecolor=blue,urlcolor=blue]{hyperref}
\usepackage{multirow}

\begin{document}

\title{Finite-temperature conductivity and magnetoconductivity of topological
insulators}

\author{Hai-Zhou Lu and Shun-Qing Shen}

\affiliation{Department of Physics, The University of Hong Kong, Pokfulam Road,
Hong Kong, China}

\date{\today }
\begin{abstract}
The electronic transport experiments on topological insulators exhibit a dilemma. A negative cusp in magnetoconductivity is widely believed as a quantum transport signature of the topological surface states, which are immune from localization and exhibit the weak antilocalization. However, the measured conductivity drops logarithmically when lowering temperature, showing a typical feature of the weak localization as in ordinary disordered metals. Here, we present a conductivity formula for massless and massive Dirac fermions as a function of magnetic field and temperature, by taking into account the electron-electron interaction and quantum interference simultaneously. The formula reconciles the dilemma by explicitly clarifying that, the temperature dependence of the conductivity is dominated by the interaction, while the magnetoconductivity is mainly contributed by the quantum interference. The theory paves the road to quantitatively study the transport in topological insulators, and can be extended to other two-dimensional Dirac-like systems, such as graphene, transition metal dichalcogenides, and silicene.
\end{abstract}
\maketitle

\begin{figure}[htbp]
\centering
\includegraphics[width=0.45\textwidth]{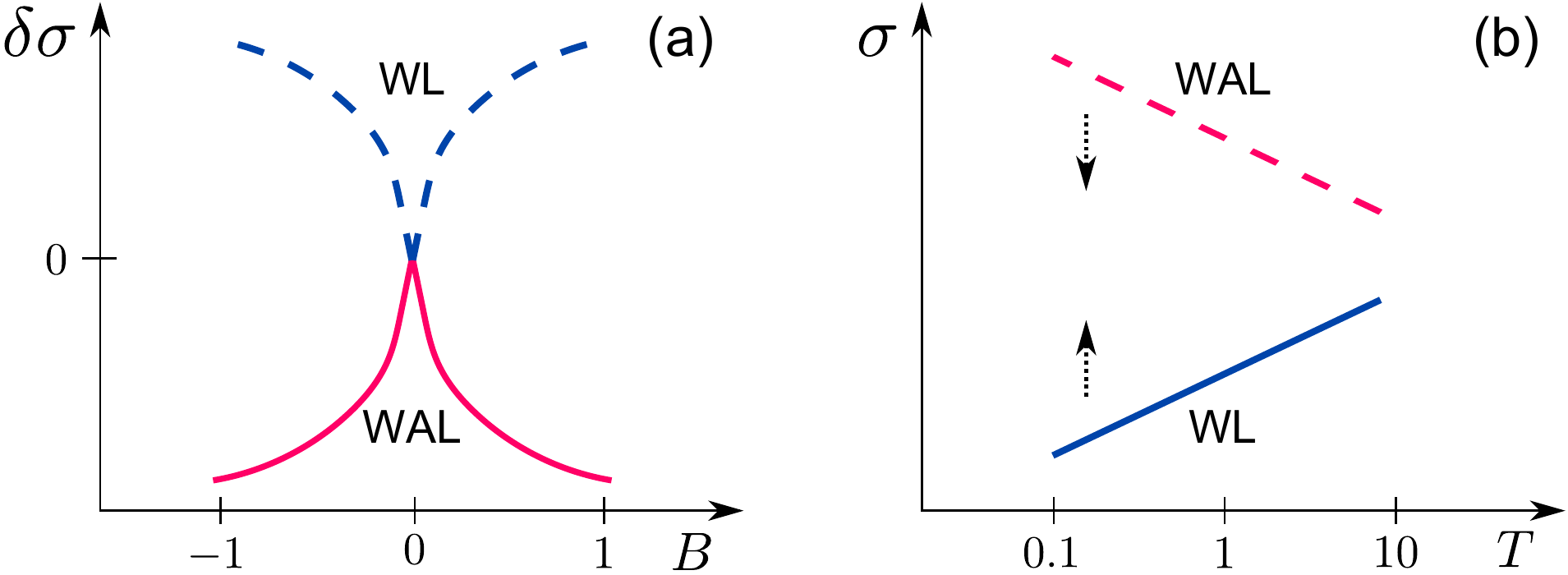}
\caption{(a) The magnetoconductivity $\delta\sigma$$\equiv$$\sigma(B)$-$\sigma(0)$, \emph{i.e.}, the change of the electrical conductivity $\sigma$ in a magnetic field $B$. (b) $\sigma$ versus temperature $T$. The dotted arrows in (b) show how $\sigma$ shifts in response to $B$, leading to $\delta\sigma$ in (a). The solid curves schematically show the contradictory $\delta\sigma(B)$ and $\sigma(T)$ measured in topological insulators: the negative $\delta\sigma(B)$ is a signature of the weak antilocalization (WAL), but the logarithmically decreasing $\sigma(T)$ with decreasing $T$ implies a tendency of the weak localization (WL).}
\label{fig:dilemma}
\end{figure}

\emph{Introduction} - The experiments on the electronic transport in topological insulators \cite{Moore10nat,Hasan10rmp,Qi11rmp,Shen12book} present a dilemma.
A negative cusp in weak-field magnetoconductivity
[solid curve in Fig. \ref{fig:dilemma}(a)] was measured in various topological insulators \cite{Checkelsky09prl,Peng10natmat,Chen10prl,Checkelsky11prl,He11prl} and commonly regarded as a signature
of the weak antilocalization (WAL) \cite{HLN80} of the surface states. For topological surface states, WAL stems from a $\pi$ Berry phase \cite{Shen04prb} acquired by electrons after circling around the spin-momentum-locked Fermi surface \cite{Xia09natphys}. When lowering temperature, the $\pi$ Berry phase induces a destructive interference between backscattered electrons, then enhances the conductivity [dashed line in Fig. \ref{fig:dilemma}(b)] \cite{Suzuura02prl,McCann06prl}. A magnetic field can destroy the interference and the conductivity enhancement, showing the negative magnetoconductivity cusp as the signature of WAL. The dilemma is, opposite to the enhancement expected from WAL, the conductivity was observed to decrease logarithmically with decreasing temperature \cite{Wang11prb,Liu11prb,Chen11prb,Takagaki12prb,Chiu13prb,Roy13apl} [solid line in Fig. \ref{fig:dilemma}(b)], indicating a behavior of the weak localization (WL) \cite{HLN80}. However, WL should exhibit a positive magnetoconductivity [dashed curve in Fig. \ref{fig:dilemma}(a)].
It has been suggested \cite{Wang11prb,Liu11prb} that the electron-electron interaction could be the possible mechanism \cite{Altshuler79ssc}, but has not been fully appreciated \cite{Bardarson13rpp},
mainly because the quantitative comparison so far \cite{Wang11prb,Liu11prb,Chen11prb,Takagaki12prb,Chiu13prb,Roy13apl} was using the theories
established for the conventional electrons \cite{Altshuler80prl,Fukuyama80jpsj,Lee85rmp}. While in topological
insulators, it is well accepted that the topological surface electrons
are massless Dirac fermions \cite{Moore10nat,Hasan10rmp,Qi11rmp,Culcer12pe,Culcer11prb,Bardarson13rpp} and the bulk electrons have to be described by a massive Dirac model to account for the topological properties properly \cite{Shen12book,Shen11SPIN}.

\begin{figure}[htbp]
\centering \includegraphics[width=0.45\textwidth]{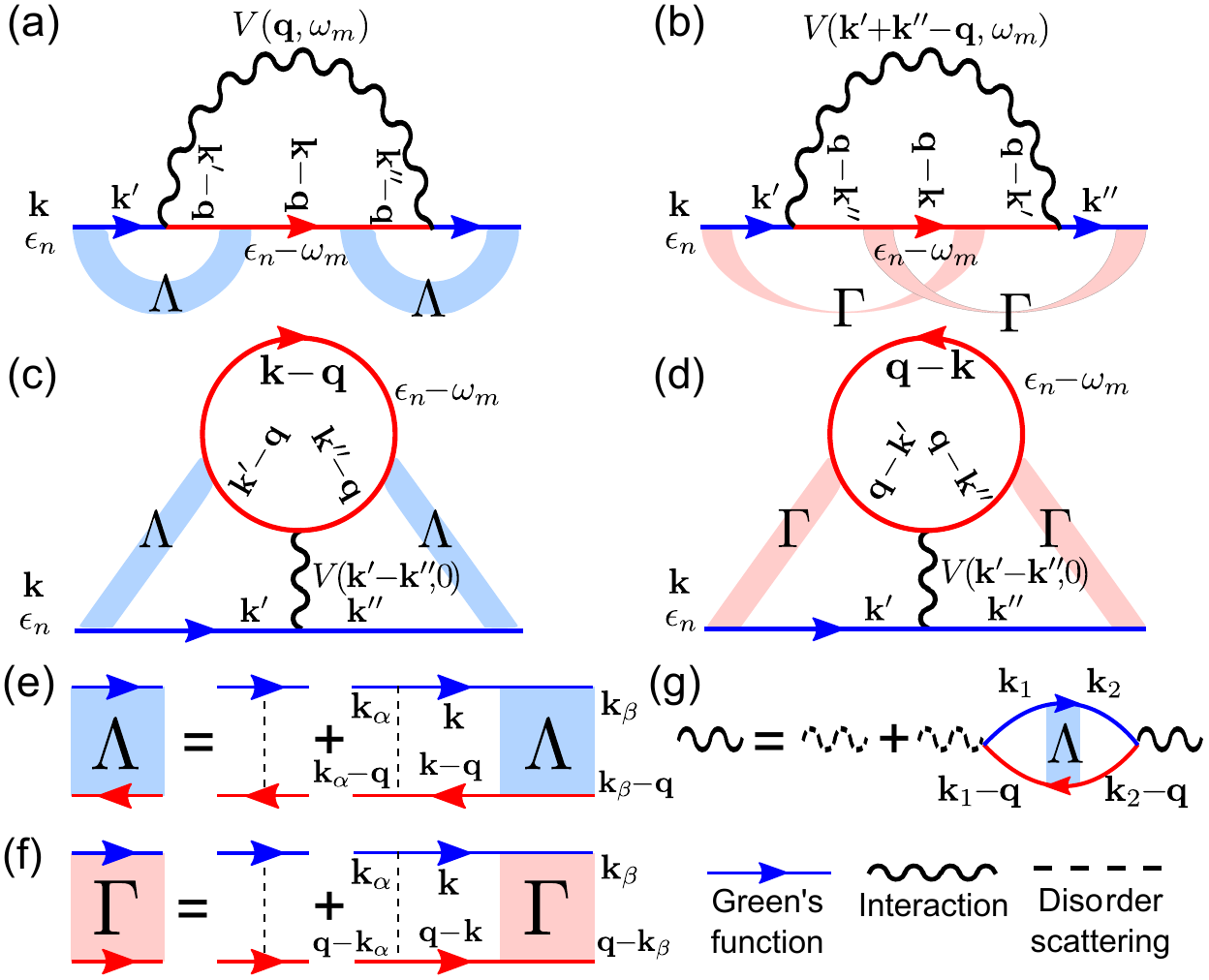}
\caption{The Feynman diagrams for the Fock [(a) and (b)] and Hartree [(c) and (d)] self-energies dressed by Diffuson (e) and Cooperon (f), from which the conductivity correction from the electron-electron interaction $\sigma^{ee}$ is calculated \cite{Altshuler79ssc,Altshuler80prl,Fukuyama80jpsj,Lee85rmp}. [(e)-(f)] The iteration equations for the Diffuson (e), Cooperon (f), and dynamically screened interaction (g). $\mathbf{k}$ and $\mathbf{q}$ stand
for the wave vectors, $\epsilon_{n}$ and $\omega_{m}$ for the Matsubara
frequencies. The diagrams for $\sigma^{qi}$ are given in Refs. \cite{McCann06prl,Lu11prl}. All diagrams can be found in Fig. S2 of \cite{Supp}.}
\label{fig:diagram}
\end{figure}

In this Letter, we resolve the dilemma by calculating the corrections to the conductivity from both the electron-electron
interaction and quantum interference for disordered massless
and massive Dirac fermions in two dimensions.
We derive a formula of conductivity as a function of temperature and magnetic field, using the diagrammatic technique (Fig. \ref{fig:diagram}). The formula reveals explicitly that in topological insulators such as Bi$_{2}$Se$_{3}$ and Bi$_2$Te$_3$: (i) The interaction always suppresses the conductivity with a strength stronger than the enhancement from the quantum interference of the surface states, leading to the WL-like temperature dependence in the conductivity. (ii) Both the interaction and quantum interference of the surface electrons produce negative magnetoconductivity, but the portion from the interaction is at least one order smaller, so the signature of WAL in magnetoconductivity mainly comes from the quantum interference. (iii) Both phenomena are attributed to a small screening factor of interaction resulting from a large permittivity in these materials. (iv) The results agree well with the experiments \cite{Wang11prb,Liu11prb,Chen11prb,Takagaki12prb,Chiu13prb,Roy13apl} at comparable temperatures (0.1 to 10 K) and magnetic fields (0 to 5 Tesla). We quantitatively compare the theory with a set of experiments for the slope of the conductivity vs. temperature, by using the Dirac mass as the
fitting parameter for both the gapless surface and gapped bulk states.
The theory is developed for massless and massive Dirac fermions, hence paves the road towards the quantitative study of the electronic transport in topological
insulators, and can be extended to other Dirac-like systems, such as graphene, transition metal dichalcogenides \cite{Novoselov05pnas,Mak10prl,Xiao12prl,Lu13prl}, and silicene \cite{Aufray10apl,Liu11prl,Chen12prl} after intervalley scattering and interaction are taken into account.

\begin{figure}[htbp]
\includegraphics[width=0.27\textwidth]{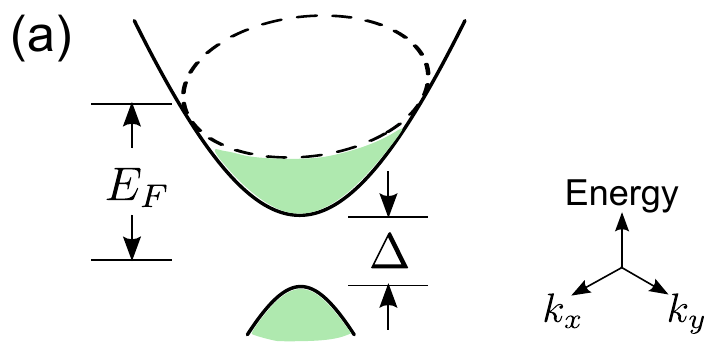}
\includegraphics[width=0.45\textwidth]{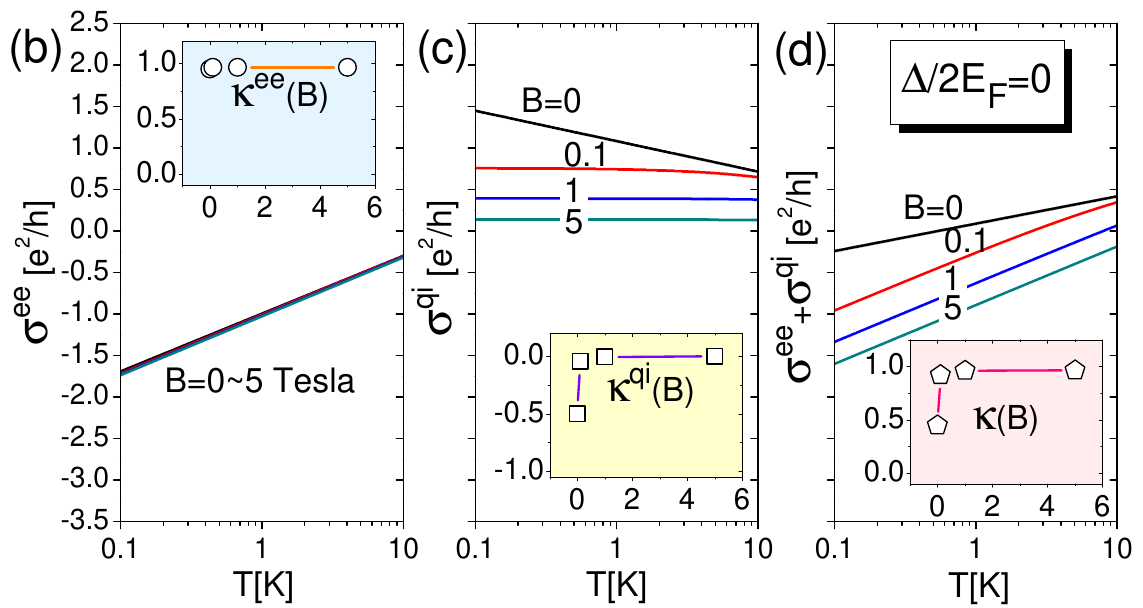}  
\includegraphics[width=0.45\textwidth]{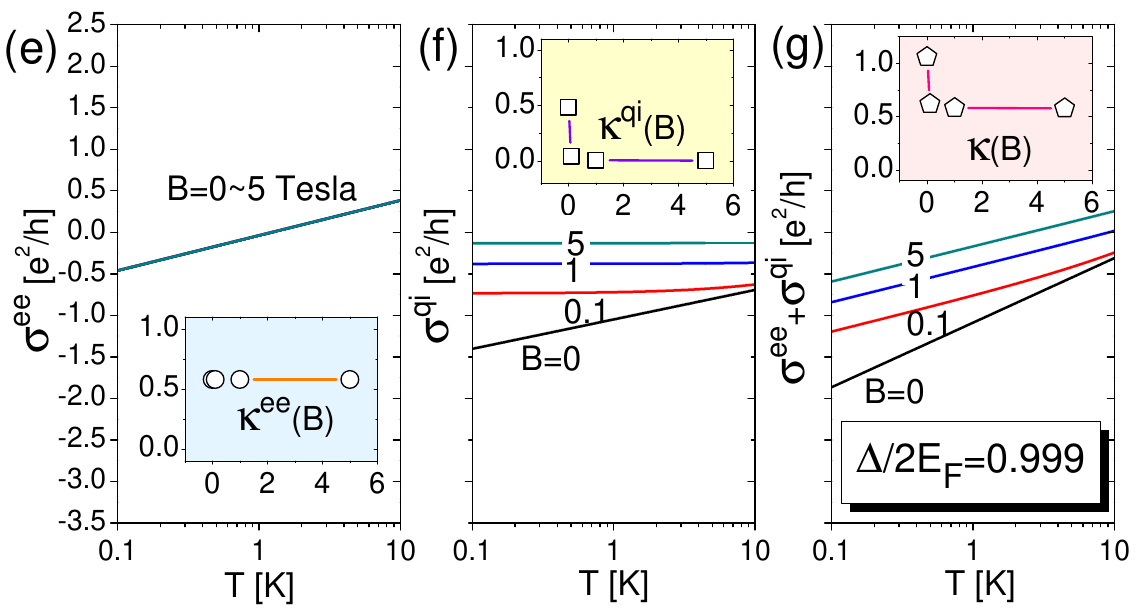}
\caption{(a) Schematic of the band structure of the 2D Dirac model. $\Delta$ is the gap (or Dirac mass). $E_F$ is the Fermi energy. [(b)-(d)] For massless Dirac fermions with $\Delta/2E_F=0$, the conductivity corrections from the electron-electron interaction $\sigma^{ee}$ and quantum interference $\sigma^{qi}$ as functions of temperature $T$ at different perpendicular magnetic fields $B$. Insets: the slope $\kappa\equiv(\pi h/e^{2})\partial\sigma/\partial\ln T$ at $T=1$ K as functions of $B$. [(e)-(g)] The same as (b)-(d) but for the large-mass limit with $\Delta/2E_{F}\rightarrow 1$, corresponding to massive Dirac fermions with $E_F$ at the band bottom. The parameters are comparable with those in Bi$_{2}$Se$_{3}$ and Bi$_{2}$Te$_{3}$: $\gamma=3$ eV\AA, the relative permittivity $\varepsilon_{r}=100$ \cite{Richter77psssb}, the mean free path $\ell=10$ nm, and the phase coherence length is taken to be $\ell_{\phi}=700\ T^{-p/2}$ nm and $p=1$ \cite{Peng10natmat,Checkelsky11prl}. The condition $\ell\ll\ell_T$ imposes a cutoff $T_H\approx \gamma/2\pi k_B \ell\approx 55$ K. The theory is valid when $T\ll T_H$, so we choose 10 K as the higher bound, consistent with the experiments \cite{Wang11prb,Liu11prb,Chen11prb,Takagaki12prb,Chiu13prb,Roy13apl}.}
\label{fig:conductivity}
\end{figure}

\emph{Model} - We start with the two-dimensional (2D) Dirac model
\begin{equation}\label{model}
H=\left[
    \begin{array}{cc}
      \Delta/2 & i\gamma (k_x-ik_y) \\
      -i\gamma (k_x+ik_y) & -\Delta/2 \\
    \end{array}
  \right],
\end{equation}
where $\gamma=v\hbar$, $v$ is the effective velocity, $\hbar$ is the reduced Planck constant, and $(k_x,k_y)$ is the wave vector. $H$ describes two energy bands with strong spin-orbit coupling, separated by a gap opened by the Dirac mass $\Delta$ [see Fig. \ref{fig:conductivity}(a)]. We assume that the Fermi energy $E_{F}$ crosses the higher band. The model has two limits: one is the massless limit with $\Delta/2E_F=0$, \emph{e.g.}, for the surface states in topological insulators; the other is the large-mass limit, which has a finite gap and the Fermi level at the band bottom such that $\Delta/2E_F\rightarrow 1$, and is applicable to the bulk electrons in topological insulator thin films near the band edges \cite{Lu11prb,Garate12prb}.

\emph{Conductivity formula} - We re-examine
the finite-temperature conductivity for 2D Dirac fermions in magnetic field. Disorder scattering and electron-electron interaction are considered when calculating the conductivity (see Sec. S2 of Ref. \cite{Supp} for details). With the help of the diagram techniques
(see Fig. \ref{fig:diagram}), we find that the temperature and magnetic field dependent conductivity can be written into two parts $\sigma=\sigma^{qi}+\sigma^{ee}$: (i) the conductivity correction from the quantum interference,
\begin{eqnarray}\label{sigma-qi}
\sigma^{qi}=\frac{e^{2}}{\pi h}\sum_{i=0,1}\alpha_{i}\left[\psi\left(1/2
+ \ell_{B}^{2}/\ell_{\phi i}^{2}\right)-\ln(\ell_{B}^{2}/\ell^{2})\right],
\end{eqnarray}
and (ii) the conductivity correction from the electron-electron interaction
\begin{eqnarray}\label{sigma-ee}
\sigma^{ee} =\frac{e^{2}}{\pi h} (1-\eta_{\Lambda\Gamma}F)\ln\frac{2\ell^{2}}{\ell_{T}^{2}}
-\frac{e^{2}}{\pi h}\eta_{\Gamma}F\psi\left(\frac{1}{2}
+\frac{\ell_{T}^{2}}{\ell_{B\phi}^{2}}\right),
\end{eqnarray}
where $e^2/h$ is the conductance quantum, $\psi$ is the digamma function, $\ell$ is the mean free path. We define $1/\ell_{\phi i}^2\equiv 1/\ell_\phi^2+1/\ell_{i}^{2}$ and $1/\ell_{B\phi }^2\equiv -( 1/2\ell_B^2 +1/\ell_{\phi 1}^2)/2\alpha_1$, where the Thouless phase coherence length $\ell_\phi$ \cite{Thouless77prl} is proportional to $ T^{-p/2}$, $T$ is temperature, and $p$ can be deduced from experiments such as Aharonov-Bohm oscillation \cite{Peng10natmat}, universal conductance fluctuation \cite{Checkelsky11prl}, and magnetoconductivity. $\ell_{B}\equiv\sqrt{\hbar/4eB}$ is the magnetic length of a perpendicular magnetic field $B$. $\ell_{T}\equiv\sqrt{D\hbar/2\pi k_{B}T}$ is the thermal diffusion length, with $k_B$ the Boltzmann constant, the diffusion coefficient $D=\ell v\sin\theta \sqrt{(1+\cos^2\theta)/(1+3\cos^2\theta)}$, and $\cos\theta\equiv \Delta/2E_F$. We find for the Dirac model, the screening factor of the interaction (see Sec. S3 of Ref. \cite{Supp} for details)
\begin{eqnarray}
F=\frac{2}{\pi}\frac{\arctan\sqrt{1/x^{2}-1}}{\sqrt{1-x^{2}}},\ \ x\equiv \frac{8\pi\varepsilon_{0}\varepsilon_{r}\gamma\sin\theta}{e^{2}},
\end{eqnarray}
where $\varepsilon_{0}$ is the vacuum permittivity, $\varepsilon_{r}$
is the relative permittivity that takes into account the effects
of the lattice ions and valence electrons. In Eqs. (\ref{sigma-qi}) and (\ref{sigma-ee}), $\ell_0^2=\ell^2 \cot^4(\theta/2)/2\alpha_0$, $\ell_1^2=-\ell^2\tan^2 \theta /4\alpha_1$,
$ \alpha_0= 4\cos^2\theta(1+\cos^2\theta)/(1+3\cos^2\theta)^2$,
$\alpha_1=-\sin^4\theta/[2(1+\cos^2\theta)(1+3\cos^2\theta)]$, $\eta_{\Lambda\Gamma}\equiv \eta_\Lambda+\eta_\Gamma$, $\eta_{\Lambda}=(1+\cos^2\theta)/2$, $\eta_{\Gamma}=-(\alpha_1/2)\sin^2\theta$. In summary, the conductivity formula is a function of temperature $T$ and magnetic field $B$, and necessarily depends on the Dirac model parameters $\Delta/2E_F$ and $\gamma$ as well as the sample-dependent parameters $\ell$, $\ell_\phi$, and $F$. The formula works in the quantum diffusion regime where $\ell\ll \ell_T \ll \ell_\phi$ and $\ell\ll\ell_B$ and when the Fermi energy is away from the Dirac point.

\begin{figure}[htbp]
\includegraphics[width=0.47\textwidth]{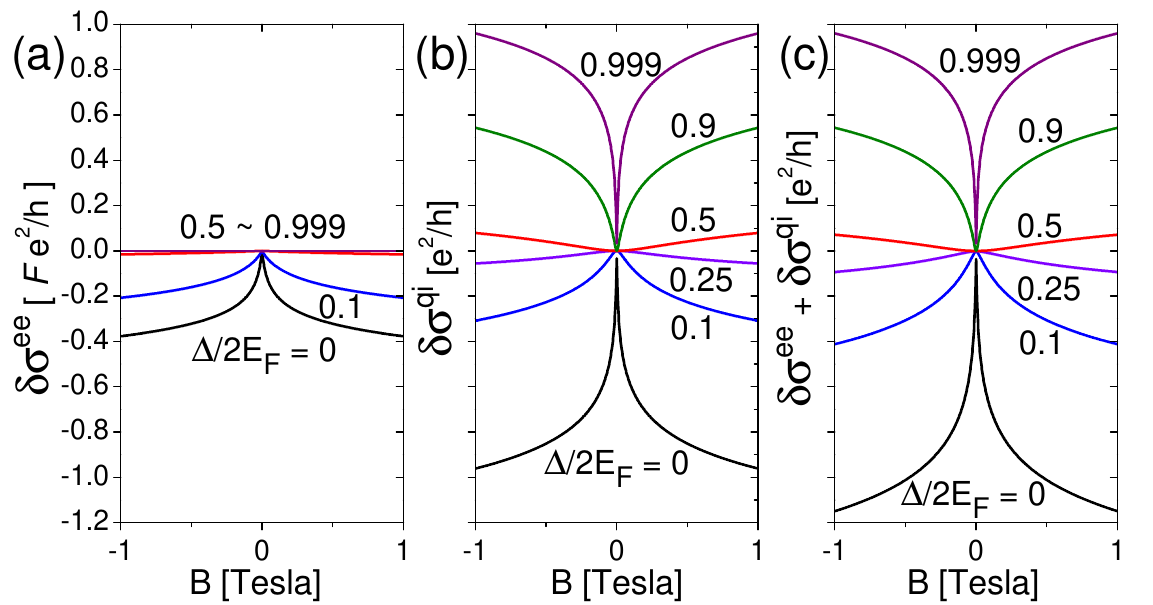} 
\includegraphics[width=0.48\textwidth]{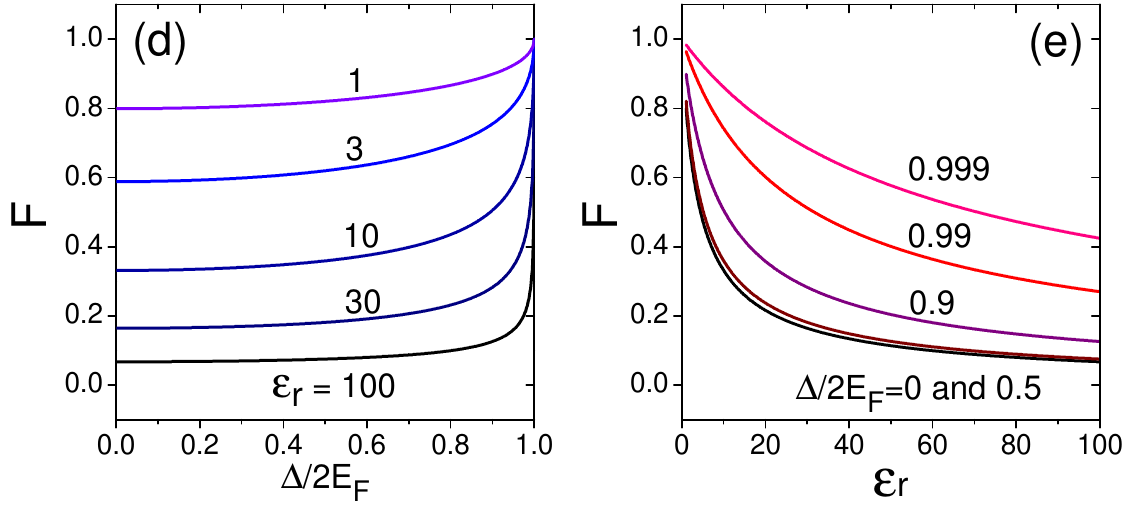} 
\caption{(a)-(c) The magnetoconductivity $\delta\sigma$$\equiv$$\sigma(B)$-$\sigma(0)$ from the electron-electron interaction $\delta\sigma^{ee}$ (in units of $Fe^2/h$) and quantum interference $\delta\sigma^{qi}$ at $T=1$ K. (d) The screening factor of interaction $F$ as a function of $\Delta/2E_F$ for different $\varepsilon_r$, the relative permittivity. (e) $F$ as a function of $\varepsilon_r$ for different $\Delta/2E_F$. The parameters are the same as those in figure \ref{fig:conductivity} except that $F=0.5$ in (c) for a better demonstration. $\varepsilon_r$ is about 100 in Bi$_2$Se$_3$ and Bi$_2$Te$_3$ \cite{Richter77psssb}.}
\label{fig:magnetoconductivity}
\end{figure}

\emph{Temperature dependence of conductivity} -
Figure \ref{fig:conductivity} compares $\sigma^{ee}$ and $\sigma^{qi}$ as functions of temperature at different magnetic fields. We first introduce the massless limit ($\Delta/2E_{F}=0$) in Figs. \ref{fig:conductivity}(b)-(d). At zero filed ($B=0$), when lowering temperature, $\sigma^{ee}$ decreases while $\sigma^{qi}$ increases logarithmically, and the total conductivity drops because of stronger $\sigma^{ee}$. The suppression of the conductivity by the interaction is also found for the leading-order conductivity at $T=0$ \cite{Culcer11prb}. The $\ln T$ behaviors can be quantitatively described by the slope
\begin{eqnarray}
\kappa\equiv(\pi h/e^{2})\partial\sigma/\partial\ln T.
\end{eqnarray}
From Eqs. (\ref{sigma-qi}) and (\ref{sigma-ee}), we found at $B=0$, the total slope $ \kappa= \alpha p+ 1-\eta_{\Lambda\Gamma}F$,
where $p=1$ from the temperature dependence of the phase coherence length $\ell_\phi\propto T^{-p/2}$ \cite{Peng10natmat,Checkelsky11prl}, and in the massless limit the Hikami prefactor \cite{HLN80} $\alpha=-1/2$ \cite{McCann06prl,Lu11prl,Tkachov11prb,Shan12prb}, $\eta_{\Lambda\Gamma}= 3/4 $ (Fig. S1 of \cite{Supp}), and the screening factor $F<0.1$ due to a large permittivity $\varepsilon_r\sim 100$ [see Figs. \ref{fig:magnetoconductivity}(d) and (e)]. Because of the small $F$, the total slope at $B=0$ in Fig. \ref{fig:conductivity}(d) is positive.
We ignore the renormalization of the Diffuson and Cooperon by the interaction, because the effect is tiny for small $F$ \cite{Finkelshtein83jetp}. Finite $B$ barely changes $\sigma^{ee}$, but suppresses $\sigma^{qi}$, giving rise to the negative magnetoconductivity at $\Delta/2E_F=0$ in Fig. \ref{fig:magnetoconductivity}(b). At large $B$, $\kappa=1-\eta_{\Lambda}F$, which increases [Fig. \ref{fig:conductivity}(d)] because the negative $\kappa^{qi}$ vanishes [Fig. \ref{fig:conductivity}(c)]. Now we move on to the large-mass limit ($\Delta/2E_{F}\rightarrow 1$) in Figs. \ref{fig:conductivity}(e)-(g). Fig. \ref{fig:conductivity}(e) shows that $\sigma^{ee}$ still decreases with decreasing $\ln T$ while its magnetic field response is completely suppressed. Figs. \ref{fig:conductivity}(f) shows that at $B=0$, $\sigma^{qi}$ becomes negative and decreases with decreasing temperature, and $\kappa^{qi}$ changes to 1/2 compared to -1/2 in Fig. \ref{fig:conductivity}(c). This change can be understood with the Berry phase $\phi_b=\pi(1-\Delta/2E_F)$ for the Dirac model \cite{Ghaemi10prl,Lu11prl}. In the massless limit, $\phi_b=\pi$, leading to a destructive quantum interference and the WAL effect \cite{Suzuura02prl,McCann06prl}. While in the large-mass limit, $\phi_b=0$, which changes the quantum interference to constructive, resulting in the crossover to the WL effect \cite{Imura09prb,Ghaemi10prl,Lu11prl} that suppresses the conductivity and reverses the sign of $\kappa^{qi}$. Fig. \ref{fig:conductivity}(f) shows that the suppressed $\sigma^{qi}$ can be recovered by the magnetic field, giving rise to the positive magnetoconductivity as $\Delta/2E_F\rightarrow 1$ in Fig. \ref{fig:magnetoconductivity}(b). Fig. \ref{fig:conductivity}(g) shows that the total slope is also positive in the large-mass limit. Between the massless and large-mass limits, $\sigma^{ee}$ and $\sigma^{qi}$ vary continuously between those in Fig. \ref{fig:conductivity}, and the total slope is always positive (Fig. S3 of \cite{Supp}). Above we show that the interaction dominates the temperature dependence of the conductivity, which always drops with decreasing $\ln T$. This agrees well with the experiments at comparable temperatures (0.1 to 10 K) and magnetic fields (0 to 5 Tesla) \cite{Wang11prb,Liu11prb,Chen11prb,Takagaki12prb,Chiu13prb,Roy13apl}.

\emph{Magnetoconductivity} - The change of the conductivity with the magnetic field in Fig. \ref{fig:conductivity} defines the magnetoconductivity $\delta\sigma$.
Figure \ref{fig:magnetoconductivity} compares
$\delta\sigma$ from the interaction and quantum interference. In the massless and large-mass limits, we find that $\delta\sigma^{qi}$ is linear in $|B|$ with $\delta\sigma^{qi}\approx -(e^{2}/ h)(2\alpha e\ell_{\phi}^2/\pi\hbar)|B|$ as $B\rightarrow 0$ or $T\rightarrow \infty$ and evolves to $\propto\ln |B|$ as $B\rightarrow \infty$ or $T\rightarrow 0$ (Sec. S1E of Ref. \cite{Supp}), consistent with the experiment \cite{He11prl}.
Fig. \ref{fig:magnetoconductivity}(b) shows that $\delta \sigma^{qi}$ changes from negative to positive as $\Delta/2E_F$ changes from 0 to 1, giving the signature of the WAL-WL crossover \cite{Lu11prl,Liu12prl,Zhang12prb,Lang13nl,Yang13prb}. In contrast, Fig. \ref{fig:magnetoconductivity}(a) shows that the magnetoconductivity from the interaction $\delta \sigma^{ee}$ is always negative, and proportional to the screening factor $F$. For the massless fermions in topological insulators, $F<0.1$ [Figs. \ref{fig:magnetoconductivity}(d) and (e)] due to a large relative permittivity $\varepsilon_r$ (typically $\sim 100$ \cite{Richter77psssb}).
In the large-mass limit $F$ approaches 1, meanwhile the $F$-independent part of $\delta\sigma^{ee}$ becomes completely suppressed [Fig. \ref{fig:magnetoconductivity}(a)].
As a result, $\delta\sigma^{ee}$ is at least one order smaller than $\delta\sigma^{qi}$. Above we show that even in the presence of interaction, the negative magnetoconductivity observed in topological insulators is mainly contributed by the quantum interference.

\begin{figure}[htbp]
\centering \includegraphics[width=0.45\textwidth]{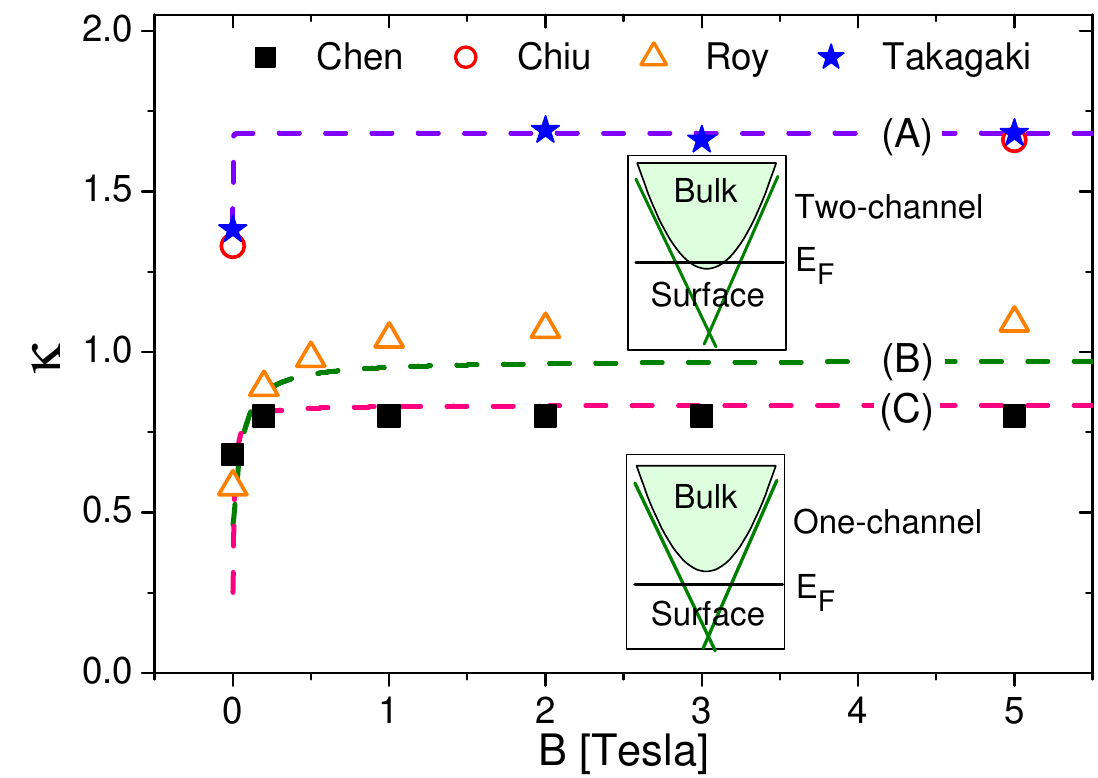}
\caption{Calculated (dashed curves) and measured (scatters) slope $\kappa$ in topological
insulators \cite{Chen11prb,Chiu13prb,Takagaki12prb,Roy13apl}. Curve
(A) is calculated from two channels by $\kappa=\kappa_{1}+\kappa_{2}$, where $\kappa_1$ is from channel 1 with $\Delta/2E_{F}=0$ for the surface band and $\kappa_2$ from channel 2 with $\Delta/2E_{F}\approx 0.98$ for the very bottom of the bulk conduction band (the higher inset).
Curves (B) and (C) are calculated from one channel with $\Delta/2E_{F}=0$, \emph{i.e.}, $E_{F}$ crosses only the surface band (the lower inset). Parameters: $\ell=10$ nm and $\gamma=3$ eV\AA. $\varepsilon_{r}=70$ and $\ell_{\phi}=700$ nm in (A); $\varepsilon_{r}=150$ and $\ell_{\phi}=130$ nm in (B); $\varepsilon_{r}=10$ and $\ell_{\phi}=300$ nm in (C). Considering varied conditions in experiments, $\varepsilon_r$ is relaxed. }
\label{fig:fitting}
\end{figure}

\emph{Slope vs. magnetic field} - The $\ln T$ and magnetic field dependence of the conductivity is characterized by the slope $\kappa$ as a function of $B$, which is summarized in Fig. \ref{fig:fitting} for recent experiments. They share two common features. (i)
$\kappa$ is always positive, consistent with Fig. \ref{fig:conductivity}. (ii) At finite $B$, $\kappa$ increases by a value of $\delta \kappa$, then saturates beyond a saturation field $B_\phi$.
The slope increase is found as $
\delta \kappa \approx -\alpha p$ for small $F$,
where $\alpha= -1/2$ at $\Delta/2E_F=0$ and $\alpha= 1/2$ as $\Delta/2E_F\rightarrow 1$ (Fig. S1 of \cite{Supp}), and $p\approx 1$ in topological insulators \cite{Peng10natmat,Checkelsky11prl}. The measured positive $\delta \kappa$ then means a negative $\alpha$ effectively, showing that massless Dirac fermions are the majority charge carriers in the experiments.
The saturation magnetic field is determined by the phase coherence length by $B_{\phi}=\hbar/4e\ell_{\phi}^{2}\approx 165/\ell_\phi^2$, with $B_\phi$ in units of Tesla and $\ell_\phi$ in nm. Most experiments have a $\ell_\phi>$ 100 nm, then $B_\phi < 0.02$ Tesla, explaining why $\kappa$ saturates after $B$ exceeds 0.1 Tesla. Also, $\kappa$ could be negative only when $B\ll B_{\phi}$, $\Delta/2E_{F}\rightarrow 0$, and $\varepsilon_r\rightarrow 1$ (Sec. S5B of Ref. \cite{Supp}).

\emph{Fitting experiments} -
As an application of our theory, we fit the experimental slopes in Fig. \ref{fig:fitting}. Take a 80 nm Cu-doped Bi$_{2}$Se$_{3}$ thin film \cite{Takagaki12prb} for example. Its sheet carrier density is about 6.7$\times$10$^{12}$/cm$^{2}$. At this carrier density, the Fermi energy is estimated to cross not only the surface band but also the very bottom of the bulk conduction band \cite{Kim12natphys} where $\Delta/2E_{F}\rightarrow1$. The experiment was originally fitted with the formula $\kappa=1-\frac{3}{4}\tilde{F}$ for the conventional electrons \cite{Lee85rmp}. However, $\kappa=1.67$ at high
$B$ yielded a negative $\tilde{F}$ which, by definition should be positive, and $\kappa$ should range between 0 and 1. Moreover, the slope change was $\delta\kappa=0.3$, differing from the theoretical value $0.5$
for a gapless Dirac cone of the surface states [inset of Fig. \ref{fig:conductivity}(b)]. The inconsistencies
imply the possibility of two channels, channel 1 with $\Delta/2E_{F}=0$
and $\delta\kappa_{1}\sim 0.5$ for the gapless surface states and
channel 2 with a large $\Delta/2E_{F}$ and $\delta\kappa_{2}\sim-0.2$
for the gapped band-edge bulk states, then $\delta\kappa=\delta\kappa_{1}+\delta\kappa_{2}=0.3$
and $\kappa=\kappa_{1}+\kappa_{2}\in(1,2)$. With $\Delta/2E_{F}$
as a fitting parameter,
the experimental slope as a function of magnetic field can be fitted by two channels very well ($\bigstar$ in Fig. \ref{fig:fitting}). Furthermore,
Fig. \ref{fig:fitting} shows that other experiments also fall in the vicinity of the $\kappa-B$ curves calculated from the Dirac model. We find that thicker films
(80 nm in \cite{Takagaki12prb} and 65 nm in \cite{Chiu13prb}) and thinner films (10 nm in \cite{Chen11prb}, 4 nm in \cite{Roy13apl}) are better fitted by two channels and one massless channel, respectively, implying the dominance of the surface states in thinner films.

We thank Michael
Ma for fruitful discussions. This work was supported by the Research Grant Council of Hong Kong under Grant No. HKU7051/11P.

\end{document}